Examining the Effects of Irradiation Temperature on Defect Generation and the Nature of Dislocation Loops

M.A. Mattucci[1], Q. Wang[1], B. Langelier[2], C. Dai[1], M. Daymond[3], N. Huin[1], C.D Judge[4]

[1]Canadian Nuclear Laboratories, 286 Plant Road, Chalk River, Ontario, Canada, K0J 1J0
[2]Canadian Centre of Electron Microscopy, McMaster University, Hamilton, Ontario, Canada
[3]Dept. of Mechanical and Materials Engineering, Queen's University, Kingston, Ontario, Canada [4]Idaho National Laboratory, Idaho Falls, Idaho, ID 83415, USA

Unlike the vast amount of irradiated material data that exists for stainless steel internals from LWRs, with high fast neutron flux and an irradiation temperature of~330$^o$C, the CANDU reactor is unique with a high thermal spectrum stainless steel components peripheral to the core. In particular, the CANDU design contains an austenitic stainless steel calandria vessel, which contains the heavy water moderator at a temperature of 60-80$^o$C. This article explores the effects of low (60-80$^o$C) and moderate (300-360$^o$C) irradiation temperature on irradiation induced defects and defect sinks, both in terms of microstructure and mechanical properties of Grade 304L stainless steel, irradiated with 3 MeV protons. State-of-the-art microscopy has been applied to characterize the irradiation defects, and nano-indentation performed to provide a link with the mechanical properties. It is hypothesized that the interstitial type of loops that develop is highly temperature dependent, and the formation of the defect loops is strongly linked with the amount of radiation induced segregation.

# 1 Introduction

Austenitic stainless steel (SS) is one of the most widely used structural materials in the nuclear industry. In LWR reactor designs, 304L, 316L and weld filler metal 308L have been extensively used for reactor internal components. For this reason, a wealth of knowledge exists in the literature pertaining to the degradation mechanisms of these materials in LWR reactor environments and there exists a rich library of ex-service material [1]–[5]. The Canadian nuclear industry finds itself at an interesting point in time, whereby many of the current fleet of CANDU reactors have been in operation for ~30 years, and with current refurbishments underway, the operational life-time of the reactor is set to extend another 30 years, with the goal of a 90-100 year life cycle. Unlike the vast amount of irradiated SS material data that exists for the LWR reactors, with high fast neutron flux and ~330$^{o}$C operating temperature, there is far less data relevant to CANDU reactor components. The CANDU design features predominantly Zr-alloy structural core components but contains a large stainless steel vessel on the core periphery, which houses the heavy water moderator. This vessel is subjected to a neutron flux with a high thermal spectrum and is in contact with water at a temperature of 60-80$^{o}$C. As a result, the operating environment is unique to the CANDU® industry.

The intent of this paper is to review the effects of low (60-80$^{o}$C), moderate (300-360$^{o}$C) and high (>360$^{o}$) irradiation temperature on irradiation induced defect size and density, in terms of microstructure and mechanical properties. In order to de-convolute the effects of irradiation temperature from other microstructural changes associated with in-reactor neutron irradiation, accelerated irradiation techniques were used to irradiate 304L SS samples at 100$^{o}$C and 360$^{o}$C. The result is an irradiation induced microstructure that is dependent on a single variable, temperature. The TEM characterization of defects and atom probe tomography (APT) analysis was previously published in [6] and the mechanical properties in [7]. The key observations from these publications that have motivated this subsequent review and discussion are;

1. The dislocation size and density distribution in the samples irradiated at 100$^{o}$C and 360$^{o}$C are comparable, with the higher irradiation temperature resulting in a larger average size and lower density, Figure 1-1. As expected, this would suggest that there is a relationship between the dislocation size and density and temperature.

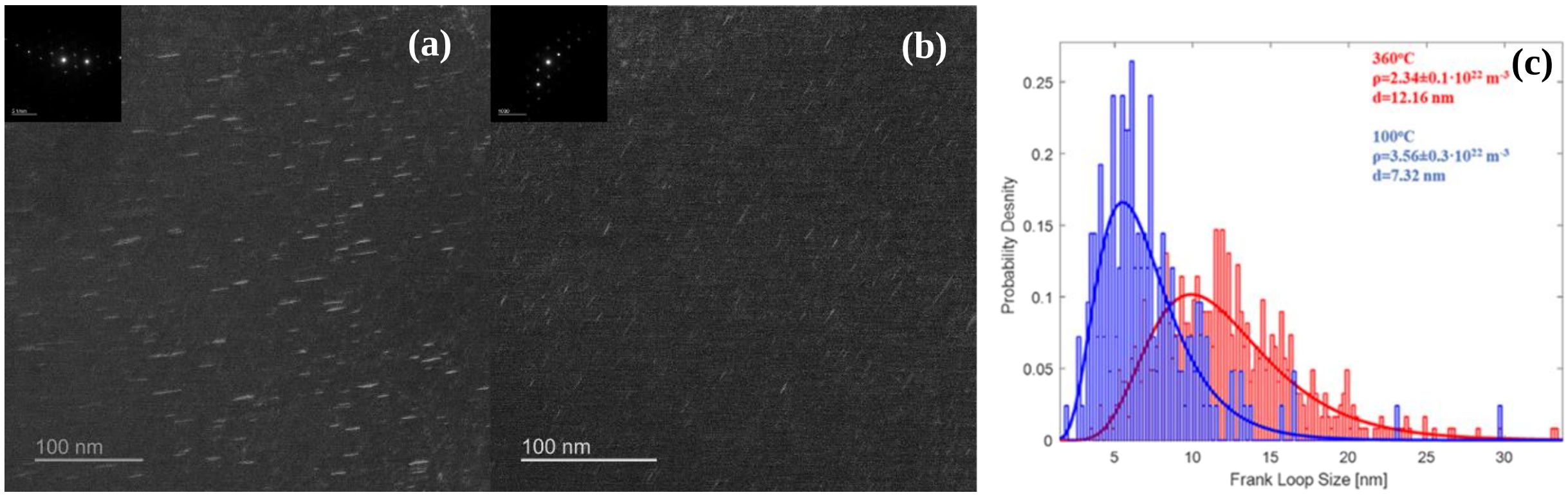

Figure 1-1: Frank loops (1/3<111>) imaged in the rel-rod condition, for (a) 360°C and (b) 100°C irradiation conditions (c) probability density as a function of Frank loop size [6].

2. APT showed visible intergranular irradiation induced segregation, i.e. enrichment of Ni and Si and depletion of Fe and Cr at dislocation loops, was observed in the sample irradiated at 360°C (Figure 1-2), while little to no segregation could be observed in the sample irradiated at 100°C (Figure 1-3). With diffusion kinetics and the mobility of species such as Ni greatly reduced at low temperature this result is as expected. However, this raises the question that if the constituents of the dislocation loops are dependent on irradiation temperature, having comparable size and density, are the properties of the individual loops dependent on the composition of the loop itself and hence also on the irradiation temperature?

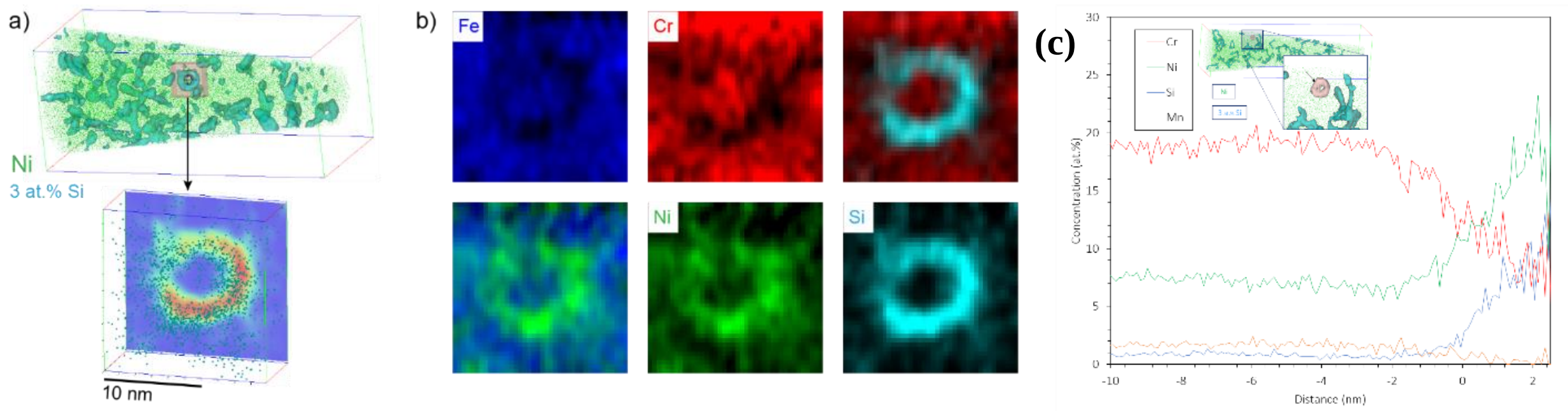

Figure 1-2: (a) Sub-volume from high temperature irradiated sample that identifies segregation to a single dislocation loop using Si isosurfaces. (b) 2D concentration of Fe, Cr, Ni and Si around the said loop, taken from the 8nm thick-sub volume. (c) A proximity histogram (proxigram) taken from the selected Si isosurface at the linear dislocation segregation region, yields the 1D concentration profiles. Results are shown for Cr, Ni, Si, and Mn, indicating Ni and Si segregation, and de-segregation of Cr and Mn. There is also de-segregation of the Fe (not shown).

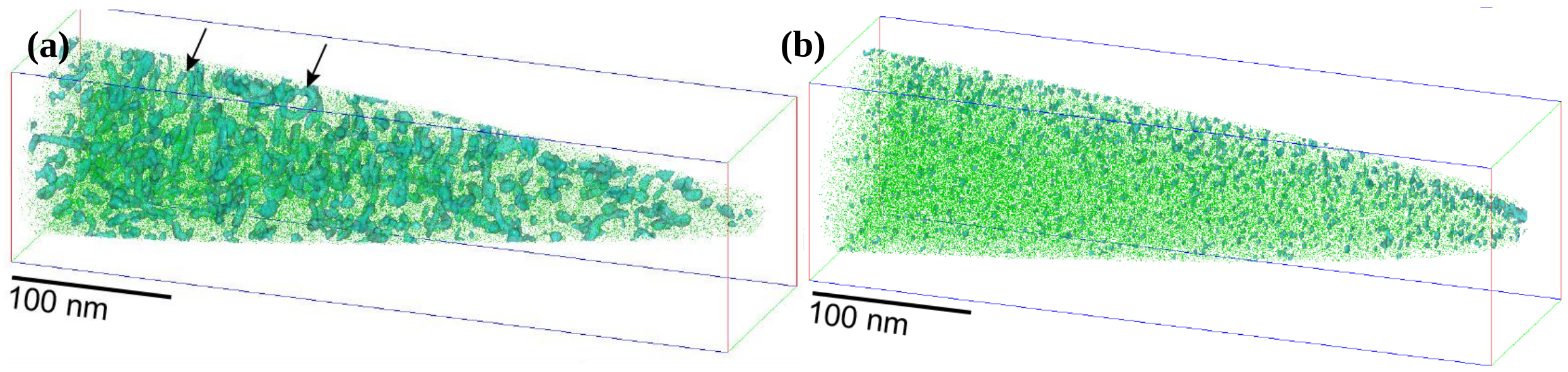

Figure 1-3: Atom maps for APT data from (a) high temperature and (b) low temperature irradiated material, shown for Ni and Si (2% and 25% of collected ions displayed for visualization purposes, respectively). Ni with isosurfaces of 2.5 at.% Si. Inset arrows indicate segregation to dislocation loops in high temperature tip.

New Chemi-STEM energy dispersive x-ray spectroscopy (EDS) data is provided to support the APT data and provide new insight into radiation induced segregation mechanisms. It is hypothesised that the nature of the irradiation defects is dependent on the irradiation temperature and that the same defects in a low temperature irradiated material will result in different mechanical properties than the analogous defects from high-temperature irradiation. This leads to the notion that considering irradiation defects, i.e., Frank loops, in the conventional sense of dislocations observed in a dark field image does not fully capture the nature of the defect.

In this review, extensions between the microstructure and mechanical properties are made, looking at the individual contribution to hardening from defect types. In addition, a review of the available literature was performed to assess the effect of irradiation temperature and dose on defect size and density [6], [8]–[20]. This review was done using available data in the literature for Frank loop size and density, dependent on dose and temperature. Tabulated data is provided in the appendix for future use. The analysis could serve as the foundation for an empirical relationship for irradiation hardening associated with Frank loops, solely dependent on irradiation dose (dpa) and temperature.

## 2 Material and Experiment

The as-received 304L plate material was hot rolled, descaled, annealed for 1 hour at 1060$^{o}$C and hot rolled again. The elemental composition is as follows: 0.019 wt.% C, 0.46 wt.% Si, 1.66 wt.% Mn, 18.19 wt.% Cr, 0.54 wt.% Mo, 0.38 wt.% Cu, 8.06 wt.% Ni, 0.129 wt.% Co, 0.032 wt.% P, 0.029 wt.% S and 0.091 wt.% N, consistent with the ASME Boiling and Pressure Vessel Code (BVPC) Section II SA-182 requirements. The grain size is categorized as 5 according to ASTM E112. The bulk hardness is 173 HB (Mechanical properties according to ASTM A370).

### 2.1 Irradiation

A series of accelerated irradiations were performed at the Michigan Ion Beam Laboratory (MIBL). The first irradiation was performed at 360±10$^{o}$C and the second irradiation at 100±10$^{o}$C. Irradiation temperatures were selected to emulate a neutron irradiation with a dose rate of $10^{-7}$ dpa/s, in a void swelling-sink dominated regime. Each irradiation was done to 1.5 dpa Kinchin-Pease (K-P), measured at 60% of the depth of the Bragg peak, calculated using the Stopping and Range of Ions in Matter (SRIM)[21] program and according to Stoller et al. [22]. The dose rate for the proton irradiation was ~$10^{-5}$ dpa/s. 2MeV protons were used to irradiate the material, resulting in a bragg peak at 18.9 µm (~21dpa K-P damage). The displacement threshold energy was set to 40 eV[22], [23] for iron-based alloys and the lattice binding energy to 0 eV.

### 2.2 Characterization

#### 2.2.1 Transmission Electron Microscopy - Defect Analysis

All TEM specimens were prepared using an FEI Versa 3D focused ion beam scanning electron-microscope (FIB-SEM) and extracted in plane-view from the plateau region (depth of 5-15 µm) and at the stopping peak (depth of 15-25 µm). The details of TEM lamella specimen preparation were previously reported in [6]. The process has been verified to remove most measurable FIB-induced defects using a non-irradiated reference specimen.

#### 2.2.2 Transmission Electron Microscopy – EDS

The EDS characterization of elemental distribution was carried out on a FEI Tecnai OSIRIS (200 KV) (scanning) transmission electron microscope (S/TEM) equipped with Bruker's XFlash detector in Reactor Materials Testing Laboratory (RMTL) at Queen's University. Every scan features a map of 1024 × 1024 pixels with the same magnification of 630kx. The spectrometric analysis was done using the Esprit 1.9 software.

### 2.2.3 Atom Probe Tomography

Specimens for APT were prepared from the damage plateau region, 5-15 µm, using FEI Versa and Zeiss NVision 40 FIB-SEM, and established lift-out techniques (e.g.[24]). Exact experimental details are reported in [6].

## 2.3 Mechanical Properties

### 2.3.1 Nanoindentation

Nanoindentation was done on the cross-section of the irradiated samples using a low load Berkovich tip and the NanoTest Vantage Indenter (Micro Materials). Using the sample cross-section, one can avoid substrate effects, a method that has been used extensively for small-scale testing on material irradiated using surface ion beam techniques [25]–[31]. The indentation size effect (ISE) was measured and the Nix-Gao model was used to measure the bulk irradiation hardness as a function of temperature [32]. These measurements were done in the irradiation plateau, close to the irradiation peak and in the non-irradiated base material, respectively. More details of sample preparation and indentation parameters are described in [7].

## 3 Results – Effects of Irradiation Temperature

### 3.1 Defect Analysis

In our previous work [6], a detailed analysis of irradiation induced defects was done. The results are summaried in Figure 3-1. With the exception of voids, the irradiation defects observed have a similar size distribution. The lower temperature material (100°C) has a smaller mean defect size, but larger number density. Recall that each sample was irradiated to 1.5 dpa and therefore the damage (cascades) induced in each sample should be equivalent, i.e., the same number of vacancy and interstitial pairs (Frenkel pairs) should be created in the damage cascade. The irradiation was performed in a sink-dominated regime and therefore net-recombination is not expected to occur, as would occur at higher irradiation temperatures, reducing the number of defects observed. This promotes the idea that there is a competition between dislocation density and size that is dependent on irradiation temperature; higher irradiation temperatures would stabilize and promote larger loop size. In order to better understand the relationship between dislocation size and density, the results were plotted in Figure 3-1 (a) and (b), respectively.

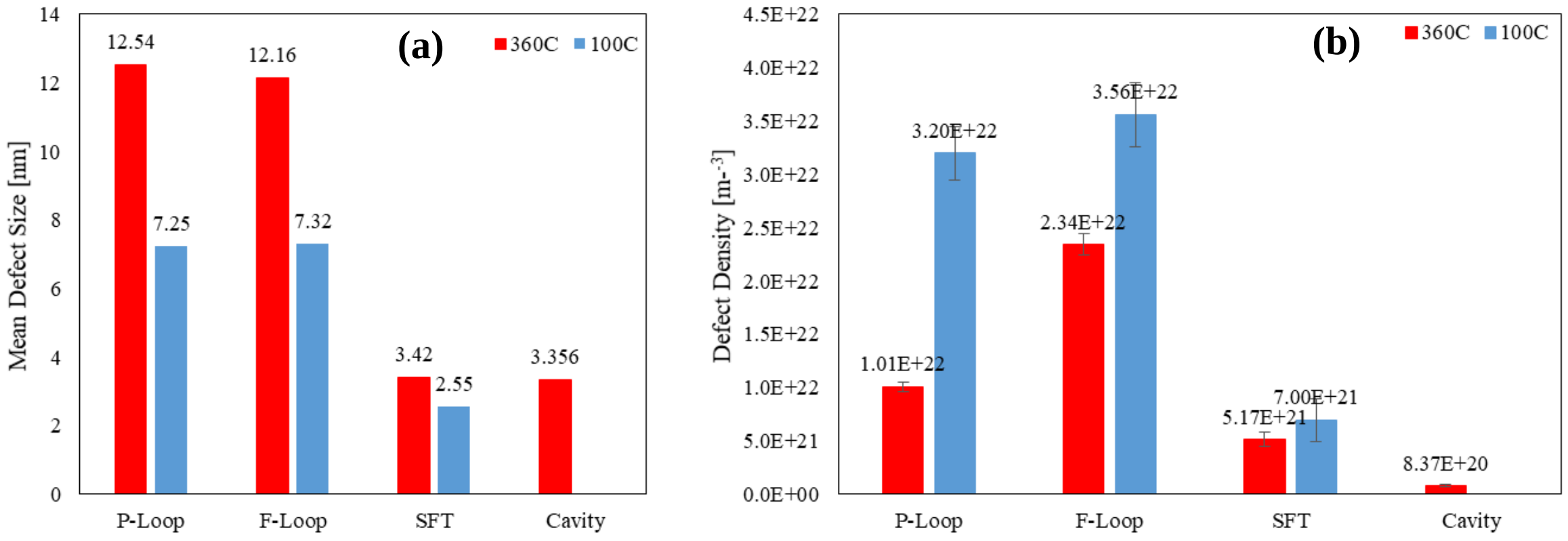

Figure 3-1: mean defect size (a) and defect density (b) of various defects. (P-loop: perfect loop, F-loop: Frank loop)

As suggested by the probability distribution function, Figure 1-1(c), the loop size increases as the density decreases. It is assumed this trend would continue for doses below irradiation damage saturation. The SFT size and density is expected to be less affected as SFT size is known to saturate between ~2 and 4 nm [33]. More on this relationship is provided in the Section 4.2.

### 3.2 Radiation Induced Segregation - Intragranular Segregation

Not only do irradiation induced defects affect the mechanical properties of the material, local chemical changes are associated with the intragranular segregation of alloying elements to and from defects [34]–[36]. Using a combination of APT and Chemi-STEM EDS the relationship of RIS to defects as a function of temperature is discussed. HAADF images and EDS scans were recorded in the grain interior for the samples irradiated at 360°C and 100°C in the 1.5 dpa plateau region and the 21 dpa damage peak region in Figure 3-2 and Figure 3-3, respectively. Note that in the damage peak, a high density of voids is likely stabilized from hydrogen deposition from the proton beam. Voids near the damage peak are shown in the HAADF image of Figure 3-3.

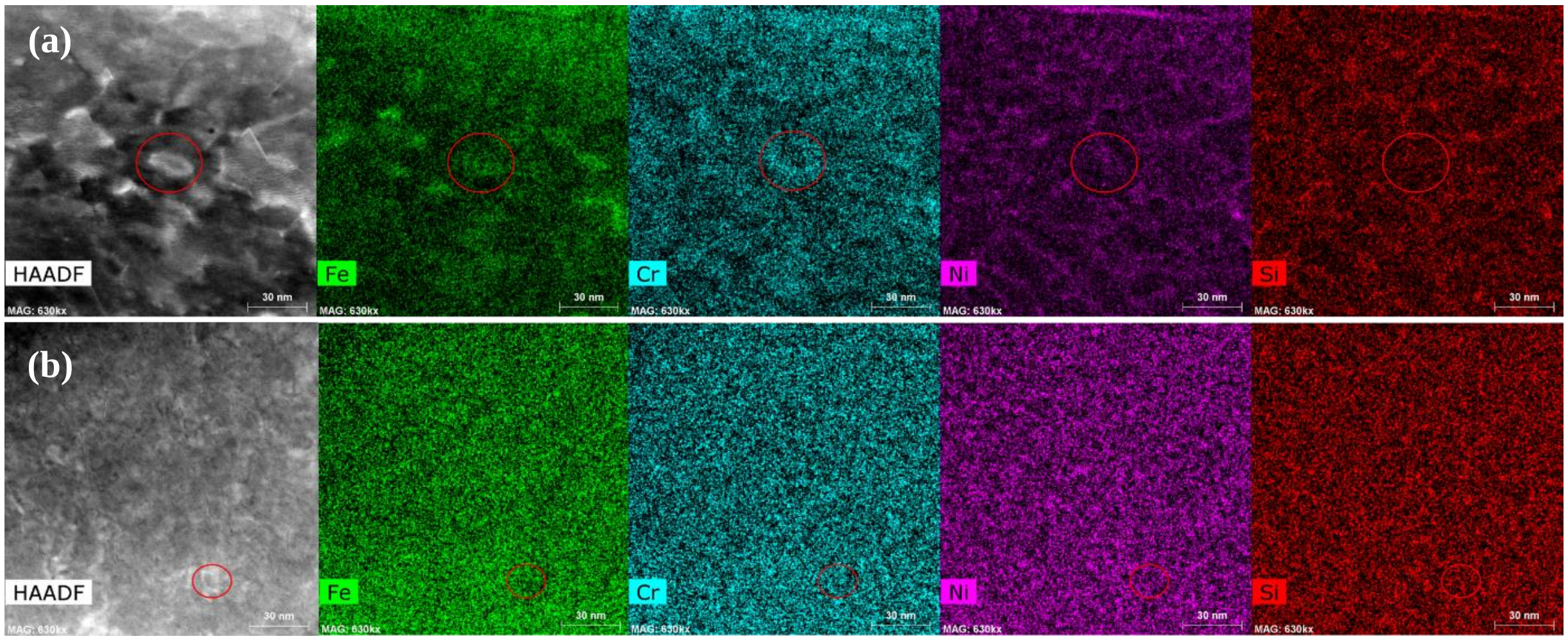

Figure 3-2: HAADF and Chemi-STEM maps in the 1.5dpa plateau region from the samples of (a) irradiated at 360 °C and (b) irradiated at 100 °C.

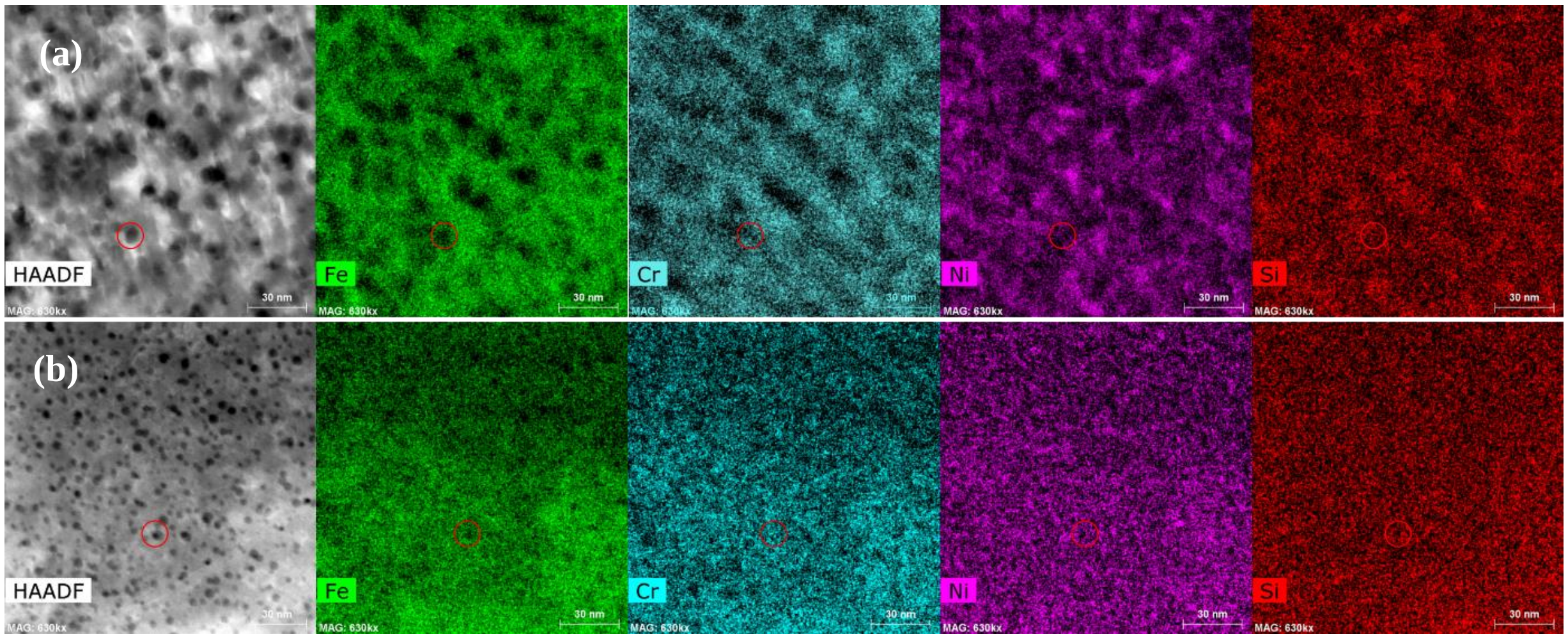

Figure 3-3: HAADF and Chemi-STEM maps in the damage peak region from the samples of (a) irradiated at 360 °C and (b) irradiated at 100 °C.

## 3.3 Intergranular Segregation

The microstructure after irradiation to 1.5 dpa at 360°C shows considerably more RIS of Fe, Cr, Ni and Si, in the grain interior than the sample irradiated at 100°C. Since RIS is a thermally driven process dependent on the radiation flux [2], the results are in line with theory. In Figure 3-2 the inset red circle on the HAADF image highlights what appears to be a dislocation loop. The same area is highlighted in the corresponding EDS images. It appears to be Ni segregation on the dislocation loop while Fe and Cr are depleted in the surrounding area of the loop. This is very similar to the common phenomenon in austenitic stainless steels after neutron irradiation where Fe and Cr are depleted from grain boundaries where Ni is enriched [3], [37]. Another point of interest is the 'nature' of the dislocation loops in high and low temperature specimens. It is unclear from post irradiation examination if these loops are formed as Ni/Si interstitial loops, or if they form as Fe interstitial loops or vacancy loops and with RIS at higher temperatures, a transition to Ni and Si enrichment at the defect occurs. These two hypotheses will be further explored in the discussion section of this paper. In the 360°C sample, clear segregation of alloy elements to the defects is observed that is however, not present in the low temperature sample (Figure 3-2 and Figure 3-3). Based on the literature [33], these dislocation loops are known to be generally interstitial in nature, but this also depends on irradiation temperature. This would suggest that the composition of the Frank loops are dependent on irradiation temperature, even though their size and distribution are comparable (Figure 1-1 (c)). The observations show that not only does the irradiation temperature impact the size and density of the loops, but it may also affect the type of interstitial atoms that form the dislocation loop. It has been shown that alloying elements can influence the nature of Frank loops (extrinsic versus intrinsic) [38] and therefore the stress state of the defect. This observation raises questions regarding whether the physical properties of the dislocation loops will be dependent not only on size and density (spacing), but also on the interstitial nature of the loop. The stress state of the dislocation loop may be dependent on the type of interstitial atoms that comprise the loop.

Figure 3-3 shows the EDS scan in the grain interior at the damage peak (~18.9 µm and 21 dpa). A large density of voids is observed in the peak as expected. Qualitatively, the size of the voids are larger, with a lower number density in the high temperature material, analogous to what was observed in the defect distributions in the plateau (1.5 dpa). The high irradiation temperature produces a lower density of larger voids, which has also been observed in other materials such as Inconel X-750, in the case of He bubbles [39].

The inset red circle on the HAADF image highlights a typical void in each material. In the corresponding EDS images, the voids are visibly free of Fe and Cr in both the low and high temperature irradiated materials. However, despite the high density of voids in the low temperature material, very little Ni segregation is observed, and this is comparable to the Ni distribution in Figure 3-2(b) at 1.5 dpa. On the contrary, the high temperature material exhibits areas enriched in and depleted of Ni, suggesting the strong interaction between Ni/Si and dislocation defects at higher temperatures. More specifically, Ni enrichment is mostly found close to the edge of large voids.

As was done with loops in the grain interior, APT measurements of grain boundary segregation were done in the 1.5 dpa regime to provide better spatial resolution. This is done by extracting samples along a grain boundary, and preparing the sample so that the boundary is positioned at the tip apex, within the volume that can be evaporated during the APT experiment. Further details can be found in [6]. Composition profiles across the boundary are obtained from sub-volume ROIs running normal to the boundary, with 30 nm diameter cross section, and are shown in Figure 3-4.

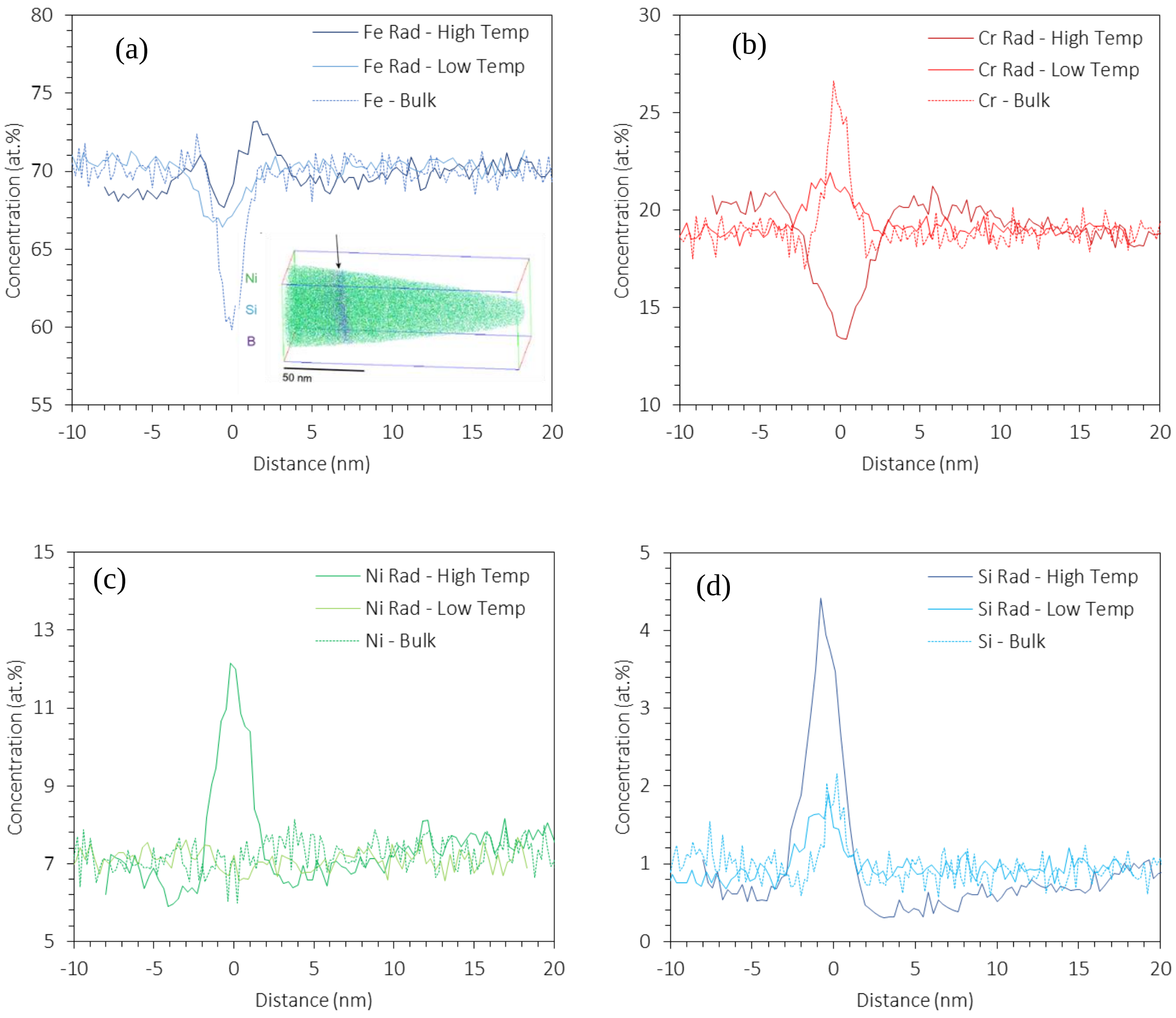

Figure 3-4: 1D concentration profile plots normal to grain boundary (located at x=0) for high temperature, low temperature irradiated and bulk (non-irradiated) datasets for (a) Fe, (b) Cr, (c) Ni and (d) Si. An example of the GB APT data set is inset in (a).

3.4 Mechanical Properties

From nano-Berkovich indentation and the Nix-Gao model, an estimate for the bulk hardness in the damage plateau, at ~5 µm and 15 µm, and in the non-irradiated material was determined. Full details of the experimental results can be found in [7]. An estimate for the Vickers hardness can be calculated from the nano-Berkovich hardness using Equation (1), which accounts for the change in indenter geometry [29], [40]:

$$H_V = 0.0945 H_{Berk} \tag{1}$$

where Vickers hardness is measured in *kg/mm²*. The results are plotted in Figure 3-5 below.

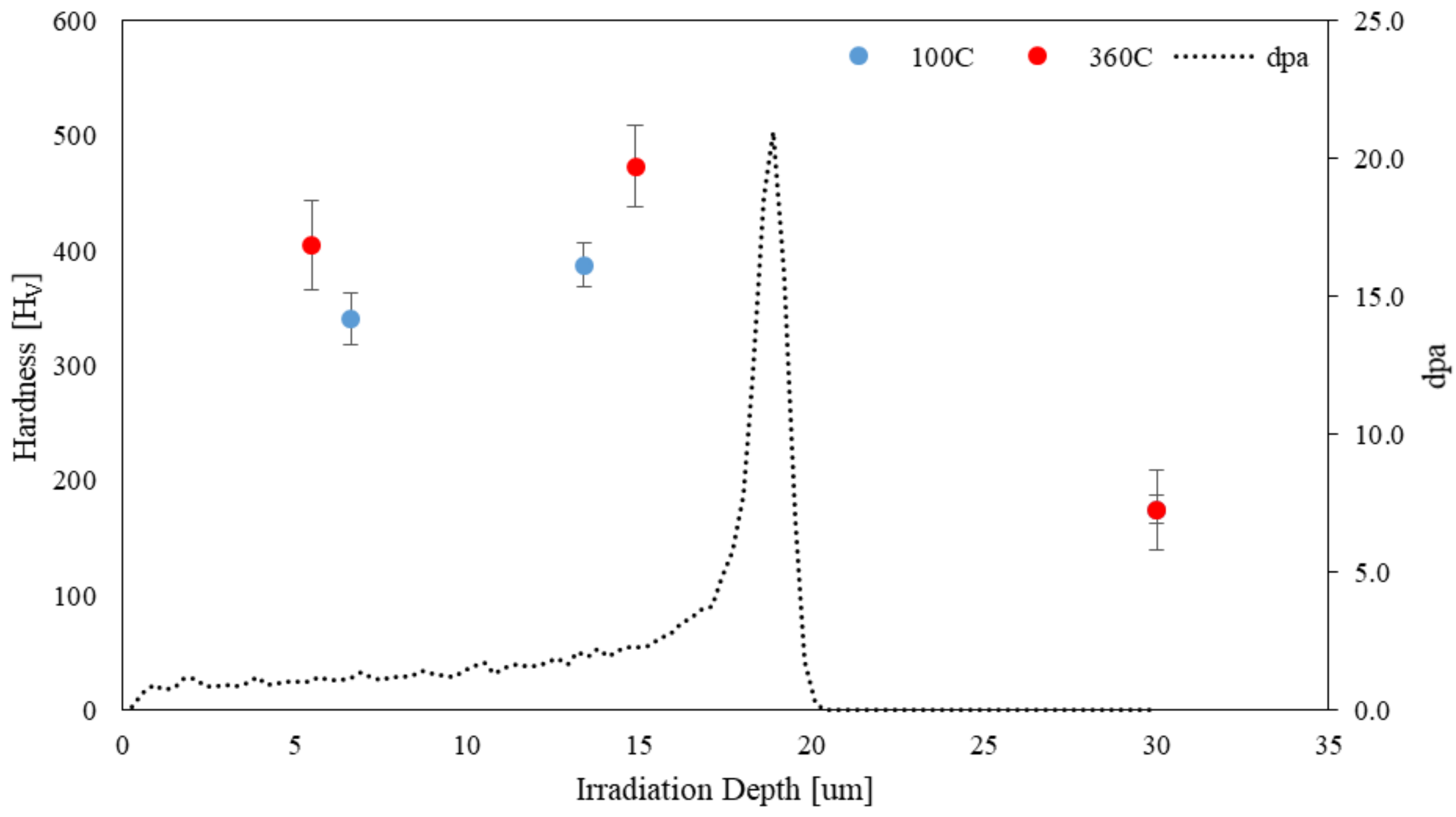

Figure 3-5: Approximate bulk Vickers's Hardness as measured from Berkovich nano-indentation as a function of irradiation depth (dpa) for the high and low temperature irradiated sample.

# 4 Discussion

## 4.1 Modelling and Mechanisms of Irradiation Induced Hardening

The empirical study done by Busby et al. [41] provides a correlation for the change in yield strength as a function of irradiation hardening, measured from micro-hardness, Equation (2):

$$\Delta\sigma_y = 3.03\Delta H_V \tag{2}$$

where Vickers hardness is measured in *kg/mm²*. From the Vickers hardness in Figure 3-5, the increase in yield from irradiation was calculated (solid bar plot in Figure 4-1(a)).

The change in yield strength can also be modelled by accounting for the obstacle hardening induced from different types of irradiation induced defects. The four types of defects characterized in this work are Perfect loops, Frank loops, SFTs and voids. We note that the resistance to deformation and interaction with obstacles varies with the nature of the defects. The relative defect barrier strength for each of the irradiation induced defects was determined from [42]. In Taylor's relationship for shear strength dependence on dislocation density, the $\alpha$ term is used to describe the defect barrier strength, on a scale from zero to one. The values obtained for irradiation induced defects are summarized in Table 1 below.

*Table 1: Obstacle strength of irraidation defects, obtained from* [42]

| Irradiation Defect | Lucas Defect | Obstacle Strength (α) |
|---|---|---|
| **Frank Loop** | Frank Loops | 0.45 |
| **Perfect Loop** | Dislocation | 0.3 |
| **SFT** | Vacancy Cluster | 0.25 |
| **Void** | Vacancy Type | 1 |

The total increase in the yield strength can be determined using the superposition principle, which accounts for the individual contribution to hardening from each defect:

$$\tau_T^n = \tau_{Frank\ loop}^n + \tau_{Perfect\ loop}^n + \tau_{SFT}^n + \tau_{Void}^n \tag{3}$$

where $\tau_T^n$ is the total increase in yield shear stress from the three defects and $n$ is a constant dependent on defect strength. For this analysis $n$ is assumed to be 2 [43], which arises from dislocation line tension approximation equations. Additionally, the yield stress can be determined from $\Delta\sigma_y = M\alpha\tau$, where $M$ is the Taylor factor, equivalent to 3.06 for FCC materials and $\alpha$ is the defect barrier strength [41].

The dispersed barrier hardening model (DBH) uses dislocation line tension approximation and considers all defects within a thin plate to interact with equal strength with the dislocation [43]. The DBH model is given by:

$$\Delta\sigma_{DBH} = M\alpha\mu b\sqrt{Nd} \tag{4}$$

where $\mu$ is the shear modulus, $b$ is the magnitude of the glide dislocation Burgers vector, $N$ is the particular defect density, measured in $nm^{-3}$ and $d$ is the average diameter of the defect. The results from the DBH model are shown in Figure 4-1(a), which agree with the values calculated from equation (2). Figure 4-1b) shows the individual contribution to hardening from each defect type. A point of interest is the similarity in hardening between defect types in the high and low temperature irradiation. The difference in hardness appears to solely arise for void/cavity formation. This supports that both defect size and distribution play an important role in hardening. This data would suggest that the smaller defect size may play a larger role in hardening than previously assumed. Equation (4) assumes that the obstacle strength is the same for each 'type' of defect and independent of irradiation temperature, namely, the hardening contribution is governed by α in Table 1. It is assumed that the obstacle strength is not effected by chemical segregation to defects or the nature of interstitial atoms of the defect.

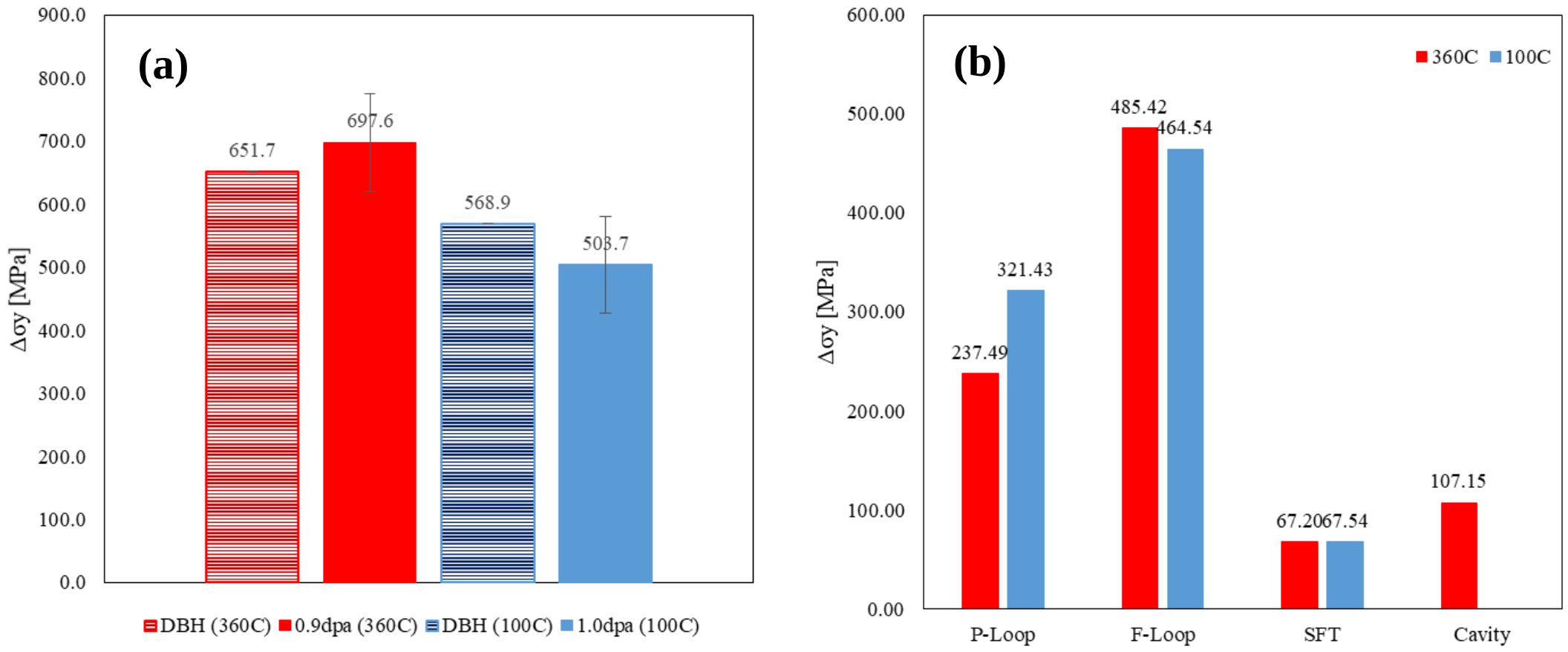

Figure 4-1: (a) Increase in yield strength as modelled from the DBH model and measured from nano-indentation for the high temperature and low temperature material and (b) the increase in yield from each type of irradiation defect as modelled from DBH model.

It should be noted that the DBH model, Figure 4-1(a), is a slight under prediction for the high temperature irradiation induced hardening and a slight over prediction for the low temperature hardening. The difference may be within error, but there is a speculative hypothesis for the difference. It could be that the obstacle strength (α) is dependent on one of two things; the size of the defects, or the alloying nature of the defect. The irradiation induced defects generated during the high temperature irradiation are both larger in size and also exhibit Ni and Si enrichment. The defects generated during the low temperature irradiation are smaller and exhibit no observable enrichment. These alloying elements may result in an additional hardening that is intrinsically incorporated in the DBH model as an over estimate of the loop strength which would therefore lead to overestimating the modelled irradiation induced hardening in the low temperature sample. This would suggest that a different obstacle strength (α in Table 3) should be used dependent on irradiation temperature. This would again support the idea that the nature of the sessile Frank loops are different in the two samples.

Another point of interest is the magnitude of the increase in yield strength. The high temperature material exhibits an increase in yield strength of 650-700 MPa, while in the low temperature material the increase is 500-550 MPa. From uniaxial tensile testing done at 288°C on the stock material, the yield strength was measured to be 167±6 MPa. Therefore the yield strength of the material irradiated to 1.5 dpa at 360°C and 100°C is about 817 and 667 MPa, respectively. Note that the nanoindentation was performed at room temperature and therefore the increase is slightly lower.

## 4.2 A Review – The Relationship between Loop Size, Density, Dose and Temperature

As discussed in Section 3.1, there appears to be a relationship between Frank loop size and density that is dependent on temperature. From the results, a shift in larger loop size appears to follow a decrease in density, and at some temperatures there is a transition in the micro-chemical composition of the interstitial loops, with the observations of Ni/Si rich interstitial loops (Figure 1-2 and Figure 1-3). The literature was reviewed extensively for tabulated values of Frank loop size and density for a given dose and irradiation temperature in both 304L and 316SS [6], [8]–[20]. The tabulated values are provided as additional information to this text. Plotting the mean loops size as a function of loop density and irradiation temperature, the relationship is not immediately clear in Figure 4-2. There appears to be some dependence of the loops' size on density and temperature.

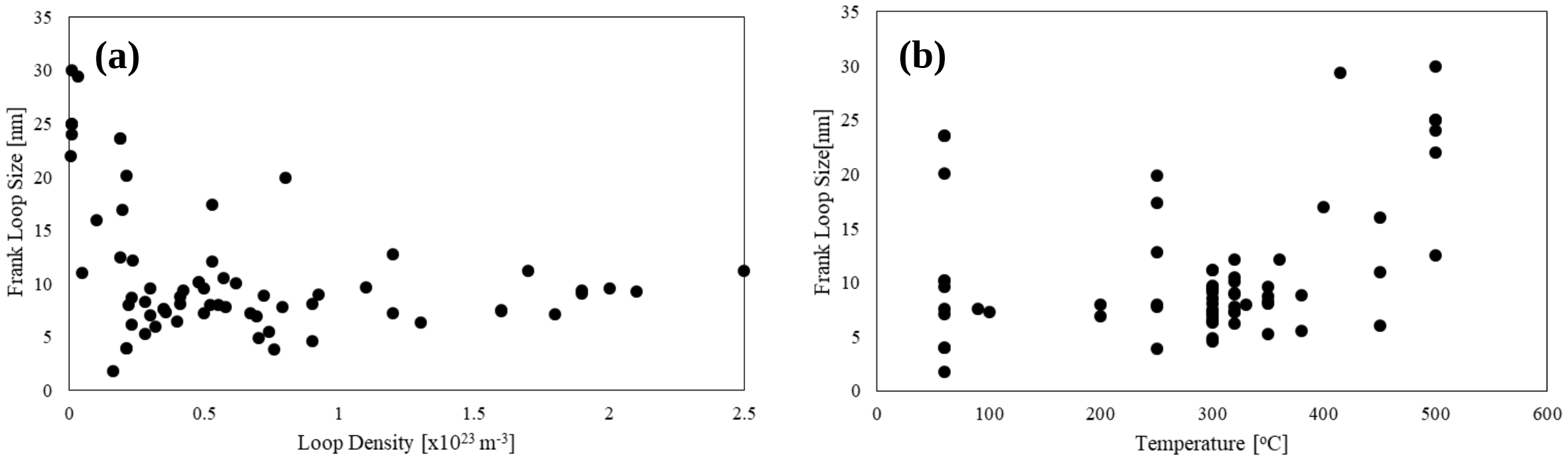

Figure 4-2:a) Frank loop size as a function of density and b) temperature with data accumulated from this study and [6], [8]–[20]

However, recall that the irradiation damage is also a function of dose and dose rate. For the purpose of this analysis we will assume the dose rates are comparable and that appropriate temperature shifts have been made during accelerated irradiations to account for the change in dose rate. The data was separated into three damage (dpa) regimes, based on when the data appeared to shift and follow a new trend, Figure 4-3. These regimes are 0-1.5 dpa, 1.5-5 dpa and >5 dpa.

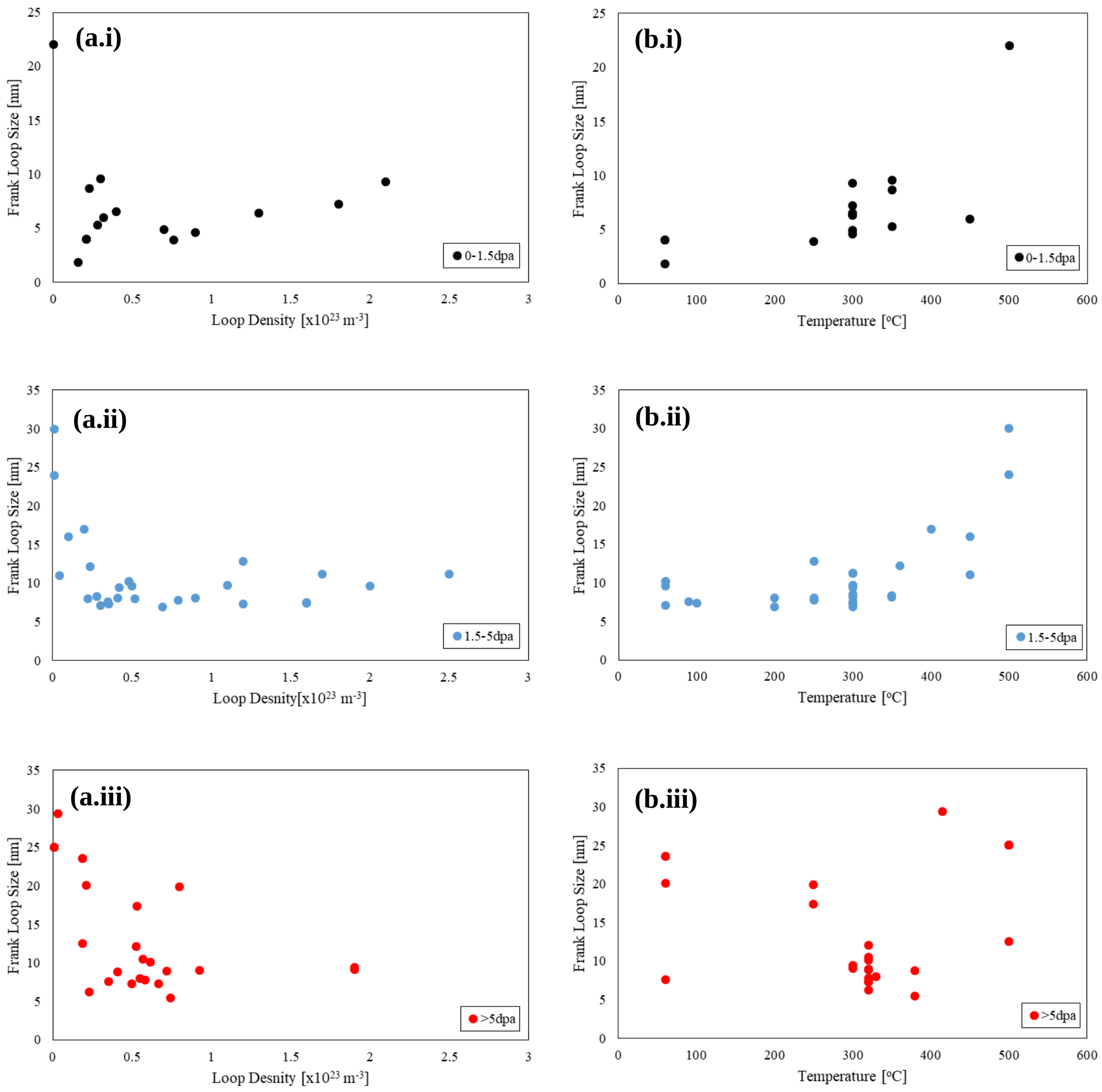

Figure 4-3: (a) Frank loop size as a function of density and (b) temperature with data accumulated from this study and [6], [8]–[20] for (i) 0-1.5dpa, (ii) 1.5-5dpa and (iii) >5dpa

Using Frank loops as an indicator for irradiation induced defect generation, the trends in the data suggest three unique regimes.

i. From 0-1.5 dpa: the onset of defect generation is occurring. The loop size is increasing as a function of loop density, which suggests the simultaneous growth and generation of defects loops. Greater than 1.5 dpa defect generation appears to saturate [44]. Additionally, the mean loop size increases with the irradiation temperature. This concept would suggest that the loops size should increase with dpa, as we know the density of defects is increasing.

ii. Between 1.5-5 dpa: appears to be the regime where defect generation stops and saturation is occurring. There is a clear relationship between the defect loop size and loop density, which suggests they are inversely proportional. A competition between the size and density is occurring, as defects compete for interstitial atoms and the formation of new loops, which is dependent on irradiation temperature. The total number of interstitial and vacancy point defects has not changed since regime i., but the distribution has. This evidence supports the claim that irradiation hardening, a function of size and density, saturates at ~5 dpa [45]. The mean loop size increases as a function of irradiation temperature, suggesting a lower density of dislocation loops.

iii. Above 5 dpa: the trend no longer exists and the loops size no longer appears to be a function of the density or the irradiation temperature. It appears that above 5 dpa loops are able to grow to a larger average size, even at low temperature.

For regimes i and ii. the maximum defect size seems to be dependent on the irradiation temperature, which also suggests that large dislocation loops >15 nm are not stable below 400$^{o}$C for irradiation doses less than 5 dpa. It is important to note this when examining loop size as a function of dpa, Figure 4-4. The tabulated data extends to 140 dpa, but was truncated after 10.5 dpa to examine the relationship in more detail up to 10.5 dpa. There is also no clear relationship beyond ~10 dpa.

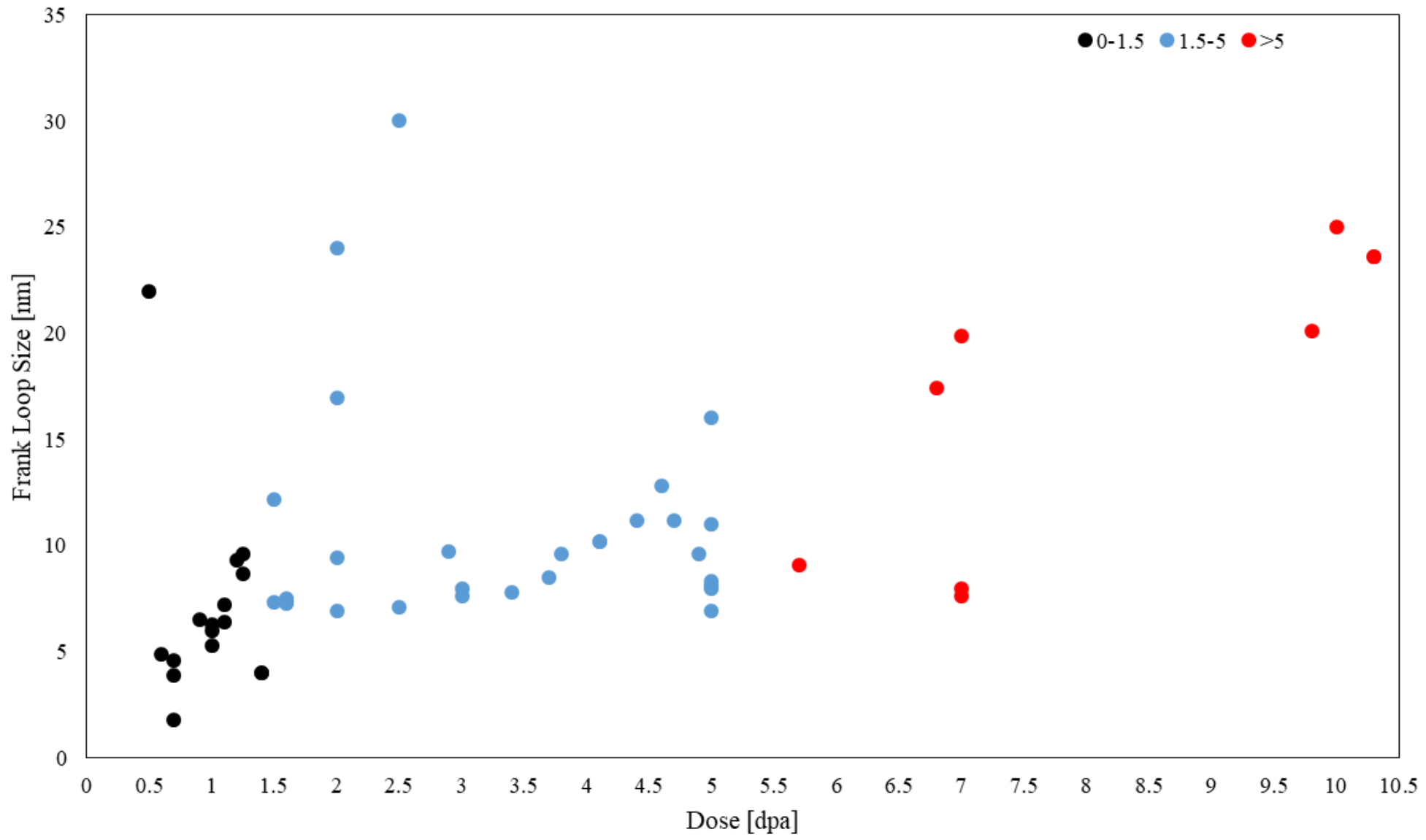

Figure 4-4: Frank loop size as a function of dpa sorted in the three dpa regimes. The data has been truncated at 10.5dpa for clarity.

As mentioned, the loop size appears to generally increase linearly from 0 to1.5 dpa. Between 1.5 and 5 dpa the loop size is relatively constant, which appears to overcome some ‘barriers’ before regime iii, the onset of saturation. It should be noted that temperature effects are not considered in Figure 4-4. In order to better understand the trends in this data, the effects of irradiation temperature must be deconvoluted.

In the Comprehensive Nuclear Materials text by S.J. Zinkle, the author discusses defect generation in terms of dpa and irradiation temperature, defining five distinct irradiation stages based on defect recovery stages [46]. The stages are defined as:

*Stage I*: The onset of long-range SIA migration, the recombination of Frenkel defects and long-range uncorrelated recombination of defects.

*Stage II*: The migration of small SIA clusters and SIA-impurity complexes.

*Stage III*: The onset of vacancy motion.

*Stage IV*: The migration of vacancy-impurity complexes.

*Stage V*: Thermal dissolution of sessile vacancy clusters.

Zinkle disuses the effect of defect generation/microstructure for irradiation temperatures regimes based on these recovery stages. In order to deconvolute the effects of temperature from the data (Figure 4-4), the data has been separated into three temperature regimes based on the description by Zinkle and apparent trends in the data.

*Stage II*: The low temperature regime (0-100$^{o}$C) - mobile SIAs, immobile vacancies.

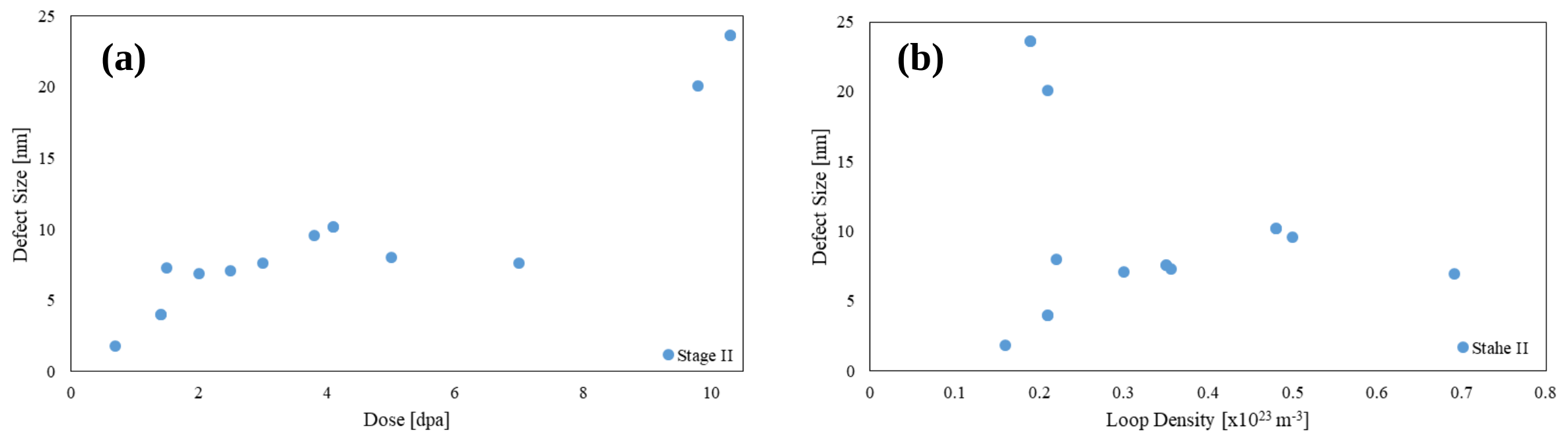

Figure 4-5: (a) Defect size as a function of dose (dpa) and (b) defect size as a function of loop density for irradiation temperature *Stage II*

In this regime SIA point defects and small SIA clusters have sufficient mobility to migrate and form visible dislocation loops and can recombine with sessile mono vacancies and vacancy clusters (SFTs) [46]. As described by Zinkle, the defect accumulation is initially linear with dose when defect concentration is low and uncorrelated recombination can occur. The relationship transitions to a square root dependence at an intermediate dose. This supports the observations in dpa regime II in Figure 4-4. Bruemmer et al. state that at this temperature, slow vacancy migration and slow vacancy emission prevent the growth of large vacancy aggregates and promote interstitial clusters. Interstitial loop growth is supressed and saturates with increasing dose [47]. Saturation in loop size occurs when the intestinal loops become dominant defect sinks and vacancy and interstitial defects annihilate at the loop at the same rate [47]. For lower irradiation temperatures saturation appears to occur at a larger fluence (dpa), when comparing Figure 4-5 and Figure 4-6 for Stage II and IV, respectively. In this regime the high sink strength of immobile vacancies limits the size of the SIA loops for lower doses (below ~5 dpa in Figure 4-5). In this temperature regime, an increase in loop size corresponds to an increase in density, Figure 4-5(b). This would suggest that at low temperatures both loop size and density increase with irradiation dose as the number of SIAs continues to increase and vacancy mobility is suppressed. Bruemmer et al. states that data suggests that, at low temperatures, the saturation of loops size is relatively insensitive to fluence/dose [3].

*Stage IV*: The medium temperature regime (100-350$^oC$) - mobile SIAs and vacancies.

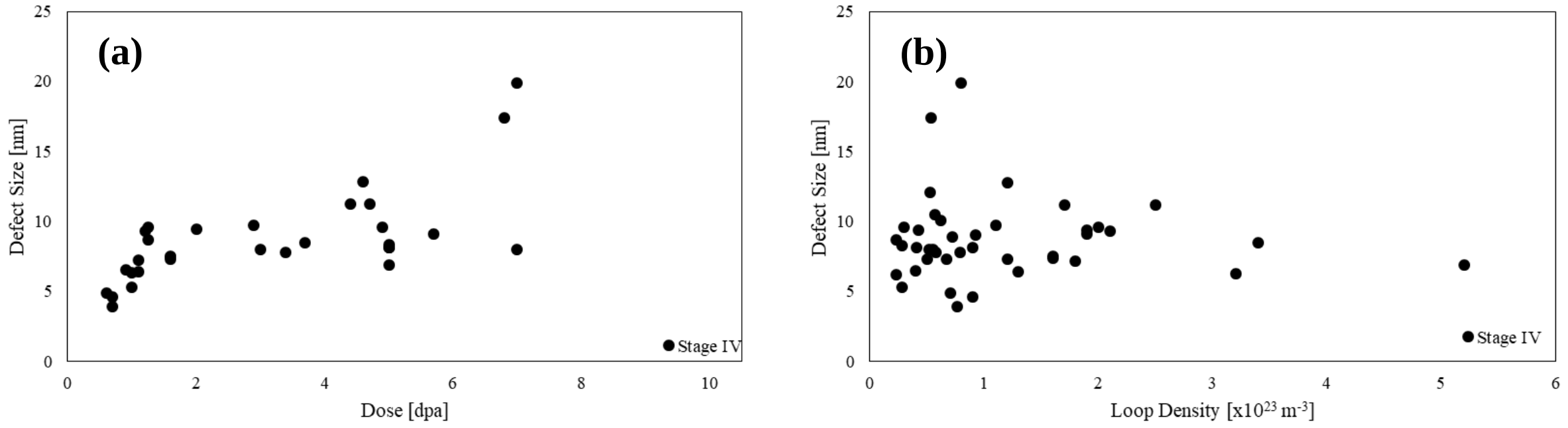

Figure 4-6 (a) Defect size as a function of dose (dpa) and (b) defect size as a function of loop density for irradiation temperature *Stage IV*

In the moderate temperature regime both SIAs and vacancies are mobile, resulting in the formation of vacancy and interstitial loops and SFTs. The majority of vacancies conglomerate in the form of SFTs and vacancy loops and as a result the majority of observed dislocation loops in FCC metals in the temperature regime are extrinsic (intestinal type) [46]. Unlike low temperature irradiations, where a high density of defect clusters suppress the saturation of vacancies as a result of recombination with point defects, near 300$^oC$ vacancy clusters become thermally stable and emit vacancies in austenitic stainless steels [47]. Microstructural evolution occurs rapidly as a result of the flux of vacancies to sinks increases [47]. Similar to temperature regime *II*, the dislocation loop accumulation is initially linear and transitions to an intermediate regime with square root kinetics, before reaching a maximum concentration level. This is directly supported in Figure 4-4 and Figure 4-6 and occurs around 5 dpa. With continued irradiation the loops can unfault and evolve into network dislocations. In this temperature regime the loop size and density does not have an apparent relationship, Figure 4-6(b). This is likely an effect of the range in dose present in the data and the size and density being more dependent on dose effects at this temperature.

*Stage V*: The high temperature regime (350-500$^oC$) – mobile defects and vacancy loop dissociation

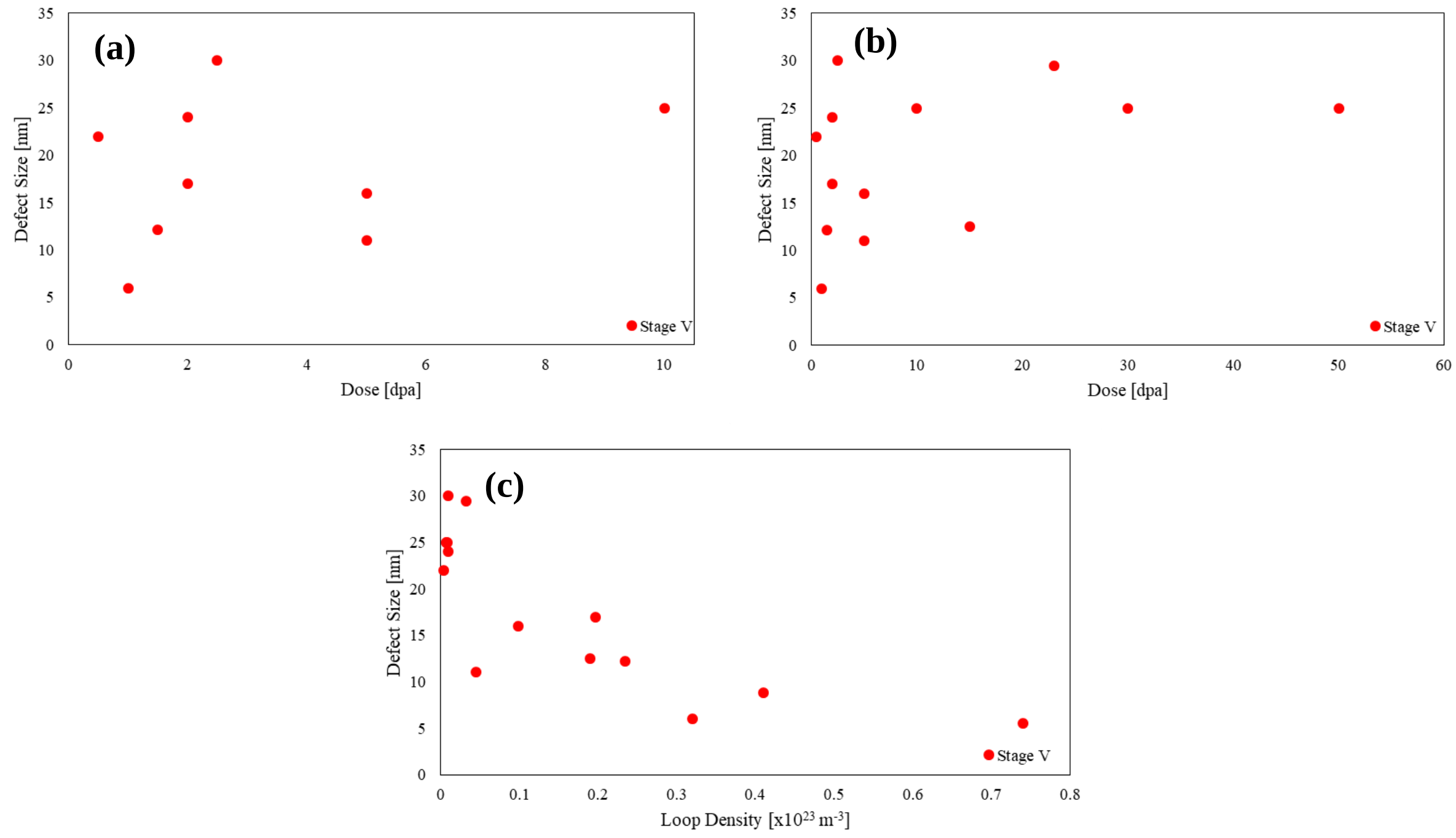

Figure 4-7: Defect size as a function of dose (dpa) for irradiation temperature *Stage V* for (a) up to 10.5 dpa and (c) up to 60 dpa and (c) defect size as a function of loops density

In the high temperature regime both vacancy and interstitial loops, network dislocation and cavities all exist in the material and SFTs are thermally unstable. In terms of defect accumulation, visible SIA clusters evolve from low density small loops to a saturation density of larger loops after damage level of 1~10 dpa. This is directly supported in Figure 4-7(b) around 10 dpa with a loop size of about 30 nm. Finally in temperature regime *Stage V*, the loop size decreases with density, Figure 4-7(c). With the increase in temperature, larger dislocation loops become stable and the preferred defect and the total density of defects decreases.

The culmination of this analysis is applying the DBH model, Section 4.1, to the wealth of Frank loop size and density data. This provides an estimate or 'window' for the increase in yield strength, from Frank loops, solely as a function of irradiation temperature. Although this may sound significant, it is noted that the variability in defect size and density for a given irradiation temperature limits the usefulness of applying the information. If anything, this observation re-affirms what is already known, i.e., irradiation at temperatures between 280$^{o}$C and 310$^{o}$C results in the largest variability in mechanical properties.

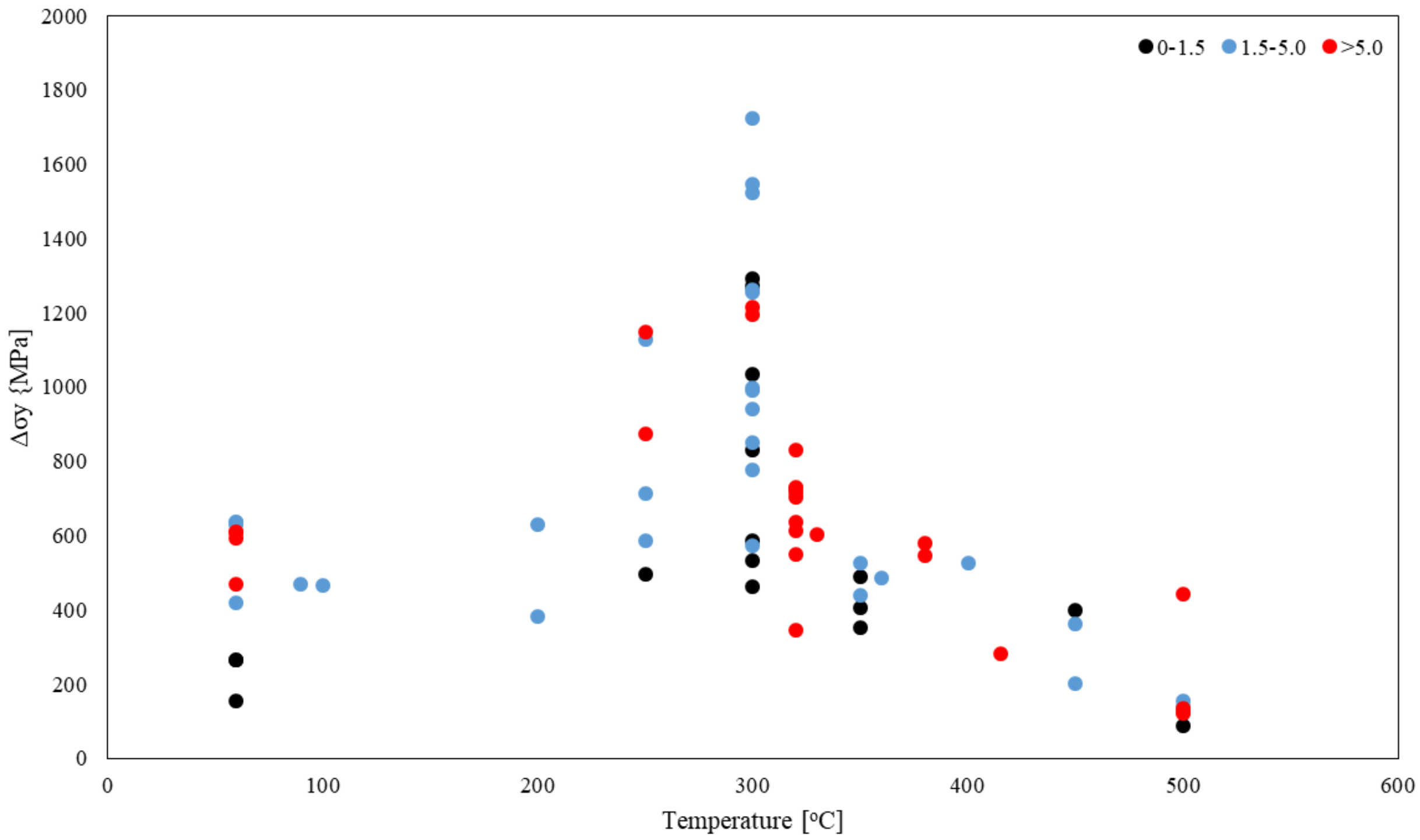

Figure 4-8: Increase in yield strength as calculated by the DBH model from Frank Loop size and density as a function of irradiation temperature.

In summary, this data provides a first approach for trying to understand the effects of irradiation temperature on defect generation, a baseline understanding of the general effect of irradiation temperature. This is a conventional approach, in-line with what is reported in the literature, which considers only the defect size and density and assumes the nature of the defects to be the same over this range. However, this study suggests it is that the very nature of the dislocation loops is also dependent on the irradiation temperature, which may affect the general properties of the loop and the material itself.

## 4.3 The Effects of Radiation Induced Segregation

In the review of stress corrosion cracking (SCC) behavior of alloys in aggressive environments by Was and Andresen, the mechanisms of RIS are discussed with particular focus on dose, dose rate and temperature. Was et al. state that the species that diffuse more slowly by a vacancy diffusion mechanism are enriched and the faster diffusion species become depleted. In addition, enrichment and depletion also occur by the association of the solute with interstitial flux [2], where undersized species enrich and oversized deplete [2], [48]. In summary RIS is dependent on several factors; the mechanism by which solutes migrate to defects, the binding energy between solutes and defects, dose, dose rate and temperature. Segregation peaks at an intermediate temperature (~300°C), due to the lack of mobility at low temperature (~100°C in this case) and back-diffusion at high temperature [2].

### 4.3.1 Dislocation Loops

Mechanically, the properties of the high and low temperature material are comparable, with additional hardening appearing to come from voids generated in the high temperature material Figure 4-1. The size distribution of irradiation defects, reported in [6] and Figure 3-1, is comparable between the low and high temperature material, with a lower density or larger defects reported in the high temperature material. As discussed, it appears to be a competition between size and density. Nevertheless, the size and distribution are on the same order. The major difference between the irradiation induced defects in high and low temperature is RIS. Fe and Cr depletion are associated with Ni and Si enrichment in dislocation loops formed during the high temperature irradiation. No noticeable segregation is observed around the dislocation loops formed during the low temperature irradiation. This agrees with the theory on the occurrence of RIS as a function of flux and irradiation temperature [2]. By this very observation, it has been concluded that the 'nature' of the dislocation loops are inherently different for the high and low temperature irradiation. Namely, at low temperature, no visible segregation is observed and the composition is homogeneous with the matrix with trace amounts of Ni and Si clustering, Figure 1-3. Based on the data in Figure 1-3 and Figure 3-2 and assuming the interstitial nature of the loops [46], [47], this would suggest that the loops formed during the low temperature irradiation are of the Fe intestinal type. To the contrary, due to intense segregation of Ni and Si observed in the high temperature irradiated sample, the stress state of the loops might be different from that of solely Fe intestinal type loops.

Based on the discussion in Section 4.2 and the review by Bruemmer and Zinkle [46], [47], it is conceivable that the mechanisms by which the loops form are temperature dependent. The formation of irradiation induced defects at low temperature (100$^{o}$C) is governed by the mobility of SIAs and the suppression of vacancy mobility, while at intermediate irradiation temperatures (360$^{o}$C) both vacancies and SIAs are readily mobile. This raises the question of whether one can simply assume the nature of dislocation loops as vacancy or interstitial type or should micro-chemical evolution also be considered when studying microstructural evolution.

It is well known that RIS is a function of the strength of defect-solute interaction and the kinetics of back diffusion (when considering grain boundaries). RIS is a non-equilibrium process that results from the flow of defects to sinks via two solute-defect interactions, Inverse Kirkendall Segregation (IKS) and Interstitial Association Segregation (IAS) [47]. Figure 4-9 shows a schematic of these interactions, taken directly from [47]. In this instance we are considering the sink as a single dislocation loops as opposed to an entire grain boundary.

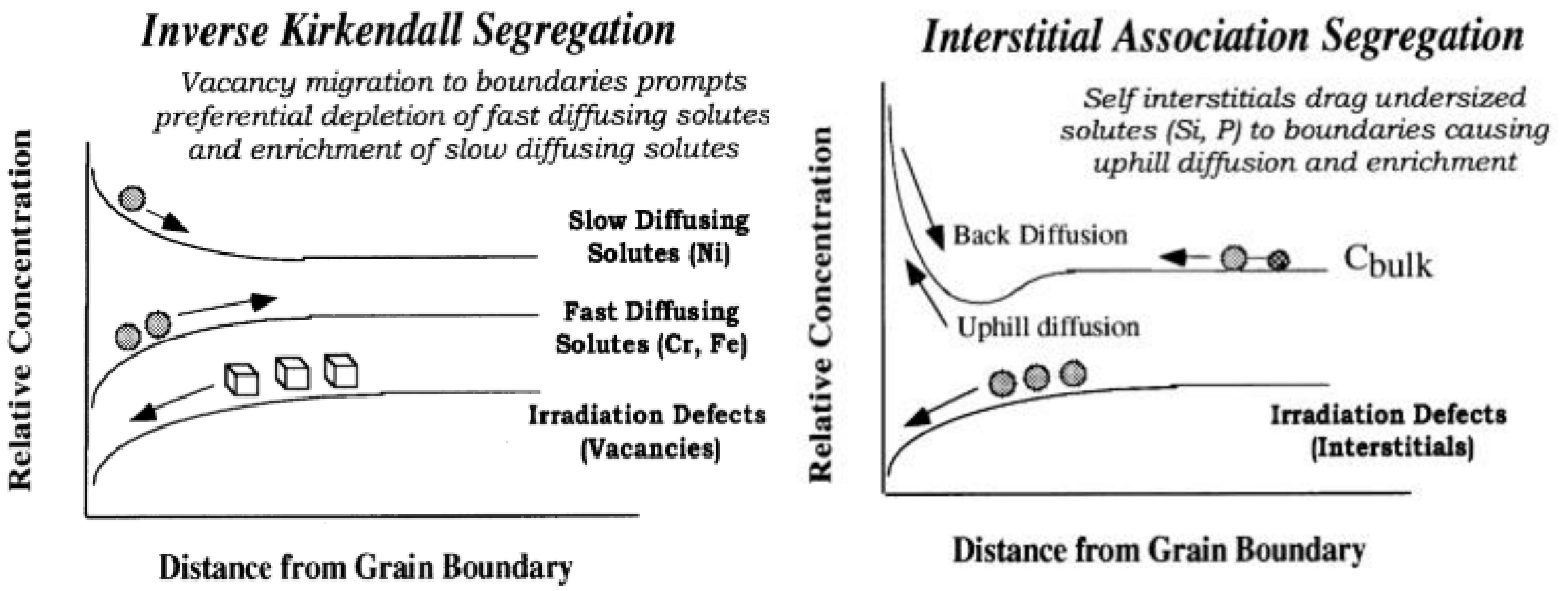

Figure 4-9: Schematic of Inverse Kirkendall Segregation and Intestinal Association Segregation taken from [47]

Bruemmer states that both mechanisms occur concurrently, but that one mechanism likely dominates. For a low temperature irradiation, *Stage II*, the vacancy mobility is greatly reduced. Therefore, it is not unreasonable to assume that IAS is dominant at low temperature and diffusion is limited to SIA mobility. Within the mobility of vacancies, chemical enrichment/depletion can be suppressed. This has been shown when modelling RIS at grain boundaries at low temperature [49]. At moderate temperature, *Stage IV*, slow diffusing elements enrich and fast diffusing elements deplete, analogous to what occurs at the grain boundaries and hence the depletion of Fe and Cr and enrichment of Ni is observed in the dislocation loops irradiated at 360$^{o}$C. The enrichment of Si at higher temperatures likely arises from an IAS mechanism whereby self-interstitials drag undersized solutes, namely Si, to the defect [47].

If we suggest that the stress state of each dislocation loop is dependent on the chemical composition or the type of interstitial atom, it is not unreasonable to assume that the matrix strength of the material may also be dependent on these micro-chemical changes. The material itself may be inherently different dependent on irradiation temperature, outside the traditional assumption that irradiation defects cause hardening. This may also suggest a difference in SCC susceptibility between the two materials. In the event of dislocation channeling to a grain boundary or crack tip [50]–[52], the different loop obstacle strength from different elements and the transport of alloying elements with dislocation motion may play a role. SCC behaviour may vary in the event of a dislocation channel when elements transported to the grain boundary or crack tip would likely be Ni and Si in the case of the high temperature irradiation, compared to Fe SIAs in the low temperature material.

# 5 Conclusion

Irradiation temperature is one of the most fundamental parameters for material ageing and environmental degradation in nuclear environments. The following conclusions have been made regarding irradiation temperature:

i. Reviewing the available literature on Frank loop characterization, a relationship between Frank loop size and density and Frank loop size and irradiation temperature can be identified. Separating the data into 'dpa regimes' reduces the scatter and identifies three probable ranges for defect generation and accumulation. The data supports saturation of irradiation hardening around 5 dpa and identifies a range for the increase in yield strength from defect loops solely dependent on irradiation temperature.
ii. As expected, the microstructural and mechanical changes from irradiation are highly temperature dependent. Lower irradiation temperatures (100$^{o}$C) produce similar defect structures (Frank Loops, Prefect Loops and SFTs) without the ability to thermally stabilize voids. As a result, higher irradiation temperatures (360$^{o}$C) resulted in larger irradiation induced hardening. Lower irradiation temperatures appear to suppress radiation induced segregation mechanisms, which may play an important role in the nature of irradiation induced defects. It is hypothesized that the nature and stress state of Frank loops is dependent on irradiation temperature, as a result of interstitial atom composition. Recent work on the effects of micro-alloying support this hypothesis [38].

Acknowledgements:

The author would like to thank Dr, Jared Smith at Canadian Nuclear Laboratories for his insightful discussions. The author would like to thank Dr. Gary Was, Dr. Ovidiu Toader and the team at the Michigan Ion Beam laboratory at the University of Michigan for their work and assistance with the accelerated irradiations. The author would thank Dr. Fei Long at the Reactor Materials Testing Laboratory at Queen's University for his assistance with STEM-EDS. This study was funded by Atomic Energy of Canada Limited, under the auspices of the Federal Nuclear Science and Technology Program. The research was conducted at the Canadian Nuclear Laboratories

# 7 Appendix

| **Ref.** | **Temperature [$^o$C]** | **Irradiation Source /Material/Sample** | **Dose [dpa]** | **Loop Density [x$10^{23}$ $m^{-3}$]** | **Mean Size [nm]** |
|---|---|---|---|---|---|
| [8] | 300 | n 304B | 0.7 | 0.9 | 4.6 |
| | 300 | n 304B | 1.6 | 1.2 | 7.3 |
| | 300 | n 304B | 5 | 5.2 | 6.9 |
| | 300 | n 304C | 0.6 | 0.7 | 4.9 |
| | 300 | n 304C | 0.9 | 0.4 | 6.5 |
| | 300 | n 304C | 5 | 0.9 | 8.1 |
| | 300 | n 304E | 1.2 | 2.1 | 9.3 |
| | 300 | n 304E | 4.4 | 2.5 | 11.2 |
| | 300 | n 316F | 1.1 | 1.8 | 7.2 |
| | 300 | n 316F | 4.7 | 1.7 | 11.2 |
| | 300 | n 316P | 1.1 | 1.3 | 6.4 |
| | 300 | n 316P | 1.6 | 1.6 | 7.5 |
| | 300 | n 316P | 2.9 | 1.1 | 9.7 |
| | 300 | n 316P | 4.9 | 2 | 9.6 |
| | 300 | n 316K | 1 | 3.2 | 6.3 |
| | 300 | n 316K | 1.6 | 1.6 | 7.4 |
| | 300 | n 316K | 3.7 | 3.4 | 8.5 |
| | 300 | n 316K | 5.7 | 1.9 | 9.1 |
| | 300 | n 316K | 13.3 | 1.9 | 9.4 |
| [13] | 300 | 8MeV $Fe^{4+}$ | 2 | 0.42 | 9.41 |
| | 400 | 8MeV $Fe^{4+}$ | 2 | 0.197 | 16.97 |
| | 200 | 8MeV $Fe^{4+}$ | 2 | 0.692 | 6.92 |
| [6] | 360 | p 304L | 1.5 | 0.234 | 12.16 |
| | 100 | p 304L | 1.5 | 0.356 | 7.32 |
| [14] | 320 | n 304L | 46 | 0.5 | 7.3 |
| | 380 | Ni++ 304L | 130 | 0.41 | 8.8 |
| | 500 | Ni++ 304L | 15 | 0.19 | 12.5 |
| | 320 | n CW 316 | 46 | 0.23 | 6.2 |
| [15] | 380 | n CW 316 | 130 | 0.74 | 5.5 |
| [16] | 415 | n AISI 304L | 23 | 0.0322 | 29.4 |
| | 320 | n 304 SA - Low S | 40 | 0.669 | 7.3 |
| | 320 | n 304 SA-high S | 40 | 0.58 | 7.8 |
| | 320 | 304 SA | 40 | 0.57 | 10.5 |
| | 320 | HP 304L SA | 40 | 0.528 | 12.1 |
| | 320 | CF-3 cast SS | 40 | 0.615 | 10.1 |
| | 320 | CF-8 cast SS | 40 | 0.923 | 9 |
| [17] | 450 | Ni ion 316 Bulk | 1 | 0.32 | 6 |
| | 450 | Ni ion 316 Bulk | 5 | 0.099 | 16 |
| | 450 | Ni ion 316 Thin | 5 | 0.045 | 11 |
| | 200 | Ni ion 316 Bulk APT | 5 | 0.22 | 8 |
| [18] | 60 | n 316 | 7 | 0.35 | 7.6 |
| | 330 | n 316 | 7 | 0.55 | 8 |
| [19] | 90 | n 316LN | 3 | 0.35 | 7.6 |
| | 250 | n 316LN | 3 | 0.52 | 8 |
| [20] | 500 | Ni++ 316 | 0.5 | 0.0043 | 22 |
| | 500 | Ni++ 316 | 2 | 0.0094 | 24 |
| | 500 | Ni++ 316 | 10 | 0.0069 | 25 |
| | 500 | Ni++ 316 | 30 | 0.0074 | 25 |
| | 500 | Ni++ 316 | 50 | 0.0088 | 25 |
| | 500 | p 316 | 2.5 | 0.0095 | 30 |
| [9] | 250 | 800Mev P 304L | 0.7 | 0.76 | 3.9 |
| | 250 | 800Mev P 304L | 3.4 | 0.79 | 7.8 |
| | 250 | 800Mev P 304L | 4.6 | 1.2 | 12.8 |
| | 250 | 800Mev P 304L | 6.8 | 0.53 | 17.4 |
| | 250 | 800Mev P 304L | 7 | 0.8 | 19.9 |
| [10] | 320 | n 316SS | 80 | 0.72 | 8.9 |
| [11] | 60 | 800MeV p 304L | 1.4 | 0.21 | 4 |
| | 60 | 800MeV p 304L | 4.1 | 0.48 | 10.2 |
| | 60 | 800MeV p 304L | 10.3 | 0.19 | 23.6 |
| | 60 | 800MeV p 316L | 0.7 | 0.16 | 1.8 |
| | 60 | 800MeV p 316L | 3.8 | 0.5 | 9.6 |
| | 60 | 800MeV p 316L | 9.8 | 0.21 | 20.1 |
| | 60 | 800MeV p 316L | 1.4 | 0.21 | 4 |
| | 60 | 800MeV p 316L | 2.5 | 0.3 | 7.1 |
| | 60 | 800MeV p 316L | 4.1 | 0.48 | 10.2 |
| | 60 | 800MeV p 316L | 10.3 | 0.19 | 23.6 |
| [12] | 350 | 160KeV Fe SA304 | 1.25 | 0.3 | 9.6 |
| | 350 | 160KeV Fe SA305 | 5 | 0.41 | 8.1 |
| | 350 | 161 KeV Fe CW 316 | 1 | 0.28 | 5.3 |
| | 350 | 162 KeV Fe CW 316 | 1.25 | 0.23 | 8.7 |
| | 350 | 163 KeV Fe CW 316 | 5 | 0.28 | 8.3 |